\documentclass[a4paper]{jpconf}
\usepackage{graphicx,amsmath,amssymb}

\usepackage{cite}

\newcommand{\ii}{\mathrm{i}}
\let \Re \relax
\DeclareMathOperator{\Re}{Re}
\let \Im \relax
\DeclareMathOperator{\Im}{Im}

\begin{document}

\title{Analytical calculation of the Green's function and Drude weight for a correlated fermion-boson system}

\author{A Alvermann$^1$, D M Edwards$^2$ and H Fehske$^1$}

\address{$^1$Institute of Physics, Ernst-Moritz-Arndt University, Greifswald, Germany}
\address{$^2$ Department of Mathematics, Imperial College London, London SW7~2BZ, United Kingdom}

\ead{alvermann@physik.uni-greifswald.de}

\begin{abstract}
In classical Drude theory the conductivity is determined by the mass of the propagating particles and the mean free path between two scattering events.
For a quantum particle this simple picture of diffusive transport loses relevance if strong correlations dominate the particle motion. We study a situation where the propagation of a fermionic particle is possible only through creation and annihilation of local bosonic excitations.
This correlated quantum transport process is outside the Drude picture, since one cannot distinguish between free propagation and intermittent scattering.
The characterization of transport is possible using the Drude weight obtained from the f-sum rule, although its interpretation in terms of free mass and mean free path breaks down.
For the situation studied we calculate the Green's function and Drude weight using a Green's functions expansion technique, and discuss their physical meaning.
\end{abstract}

\section{Motivation}

The motion of a fermionic particle, such as an electron or hole, that interacts strongly with some
background medium is a constantly recurring theme in condensed matter
physics.
The present contribution is motivated partly by the study of the
motion of a hole in an antiferromagnetic spin background, as
in a doped Mott insulator. Mott insulators occur for
electrons on a lattice subject to strong Coulomb repulsion~\cite{Mo90},
and are commonly studied in the context of the Hubbard model.
Since putting two electrons at the same lattice site costs a large energy if
Coulomb repulsion is strong, the number of doubly occupied lattice sites is negligibly small.
At half-filling we can therefore assume that every lattice site is occupied by a single electron. 
The system is an insulator because electron motion, which creates doubly occupied sites, is energetically forbidden.

\begin{figure}
\begin{minipage}[b]{0.3\textwidth}
\centering
\includegraphics[width=0.99\textwidth]{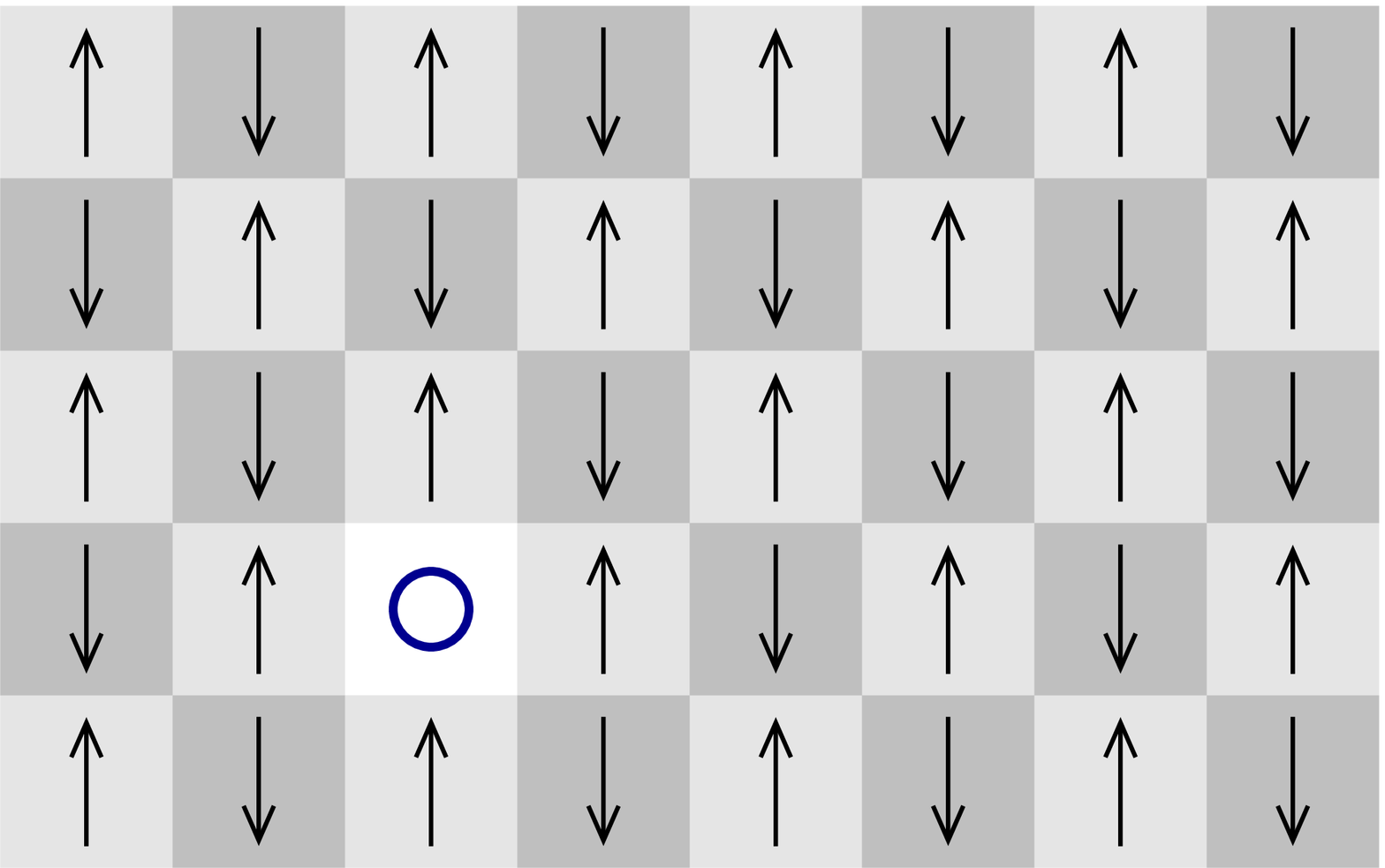}\\[1em]
\includegraphics[width=0.99\textwidth]{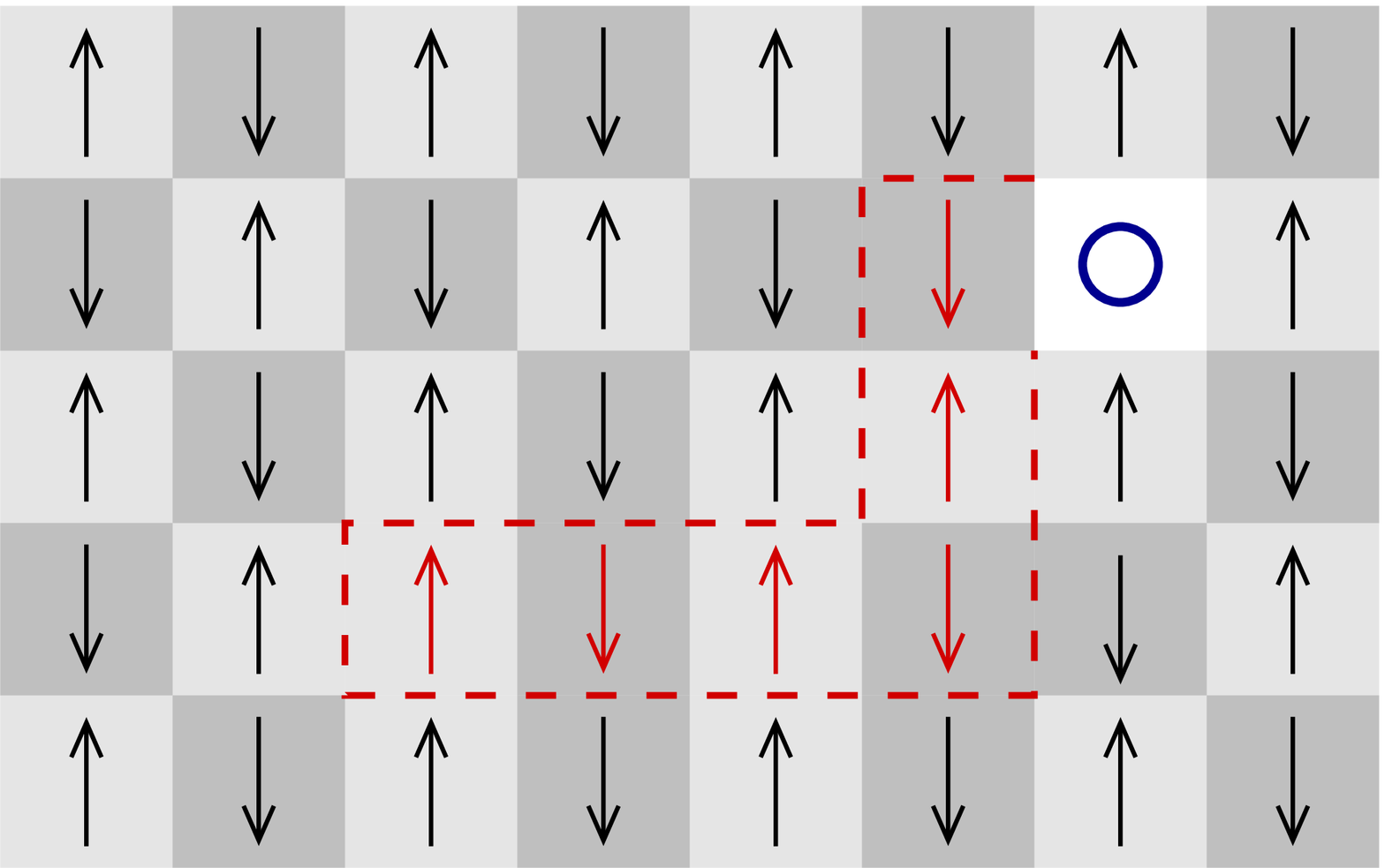}
\end{minipage}
\hfill
\begin{minipage}[b]{0.65\textwidth}
\caption{\label{fig:holeneel}
Upper panel: Hole in a N\'{e}el antiferromagnet on a square lattice.
Lower panel: When the hole moves by nearest-neighbour hopping, it creates misaligned spins that do not match the N\'{e}el order.
These distortions of the spin background lead to parallel spins on
neighbouring sites -- indicated with a dashed line -- with a substantial
increase in energy. To `unwind' the string of background distortions,
the hole in principle has to retrace the path it moved, and is therefore bound to its origin. 
The existence of (quantum) spin fluctuations, and spin-order restoring processes (see Fig.~\ref{fig:trugpath}) weakens, or completely destroys, this `string effect'. 
}
\end{minipage}
\end{figure}

If electron motion is suppressed, the electron spin is the only relevant degree of freedom. 
Perturbation theory in the kinetic energy shows that spins couple antiferromagnetically in a Mott insulator. Starting from the Hubbard model, spin dynamics is described by the antiferromagnetic Heisenberg model with complex properties arising from the interplay of strong correlations and quantum spin fluctuations.
In a very simple picture, we may alternatively consider the N\'{e}el state of a classical antiferromagnet (see Fig.~\ref{fig:holeneel}, upper panel).

If an electron is removed from a two-dimensional (2D) Mott insulator in a N\'{e}el state, the hole can move without creating doubly occupied sites.
But the moving hole distorts the antiferromagnetic spin background, creating misaligned spins with substantial energy (see Fig.~\ref{fig:holeneel}, lower panel). 
The `string' of misaligned spins that forms along the path of the hole strongly restricts propagation, and tends to bind the hole to its origin.  
Therefore the hole does not move as a free particle.
In a quantum antiferromagnet, which does not possess the strict order of the N\'{e}el antiferromagnet, the spin distortions can `heal out' or relax by quantum spin fluctuations.
It is then the relaxation rate of spin distortions that determines how fast the hole can move: The hole is continuously creating a string of distortions but can move on `freely' at a speed which gives the distortions time to decay.

\begin{figure}[b]
\centering
\includegraphics[width=0.13\textwidth]{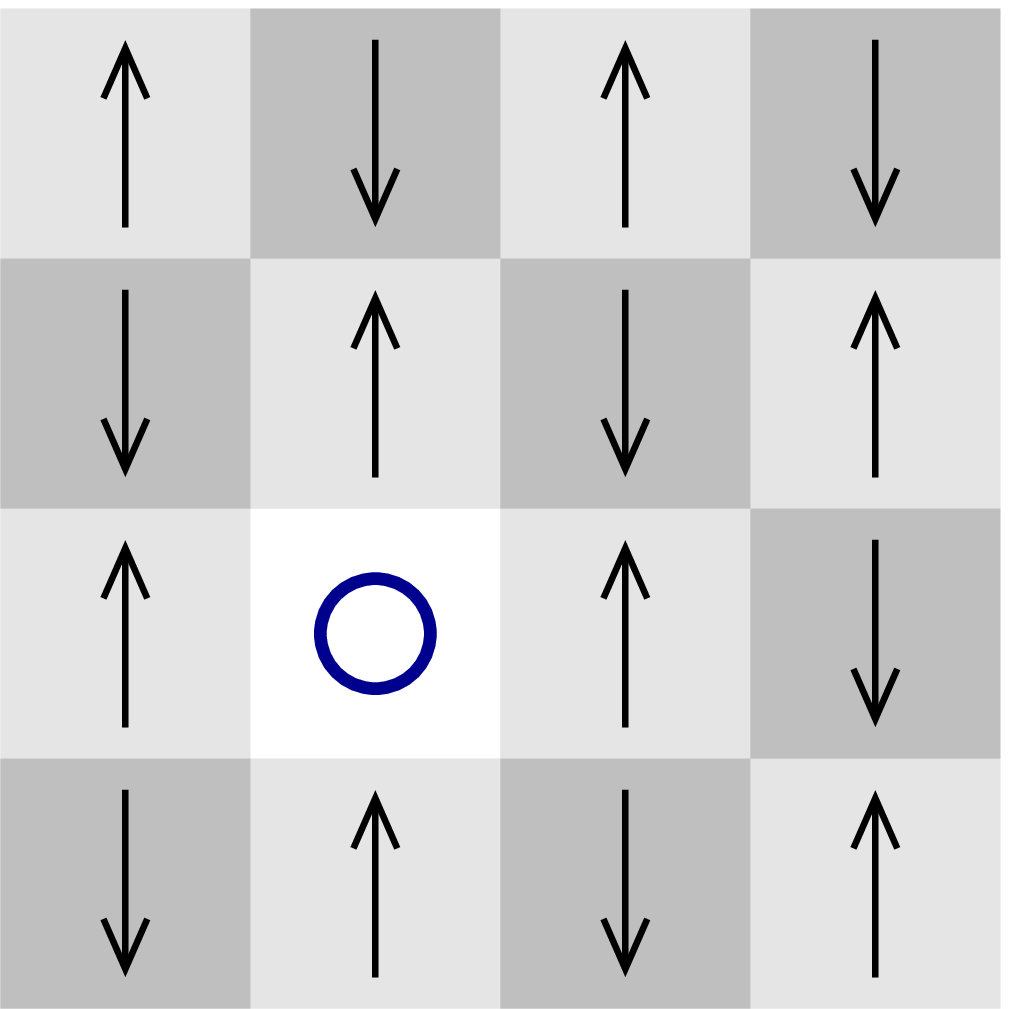}\hspace{0.005\textwidth}
\includegraphics[width=0.13\textwidth]{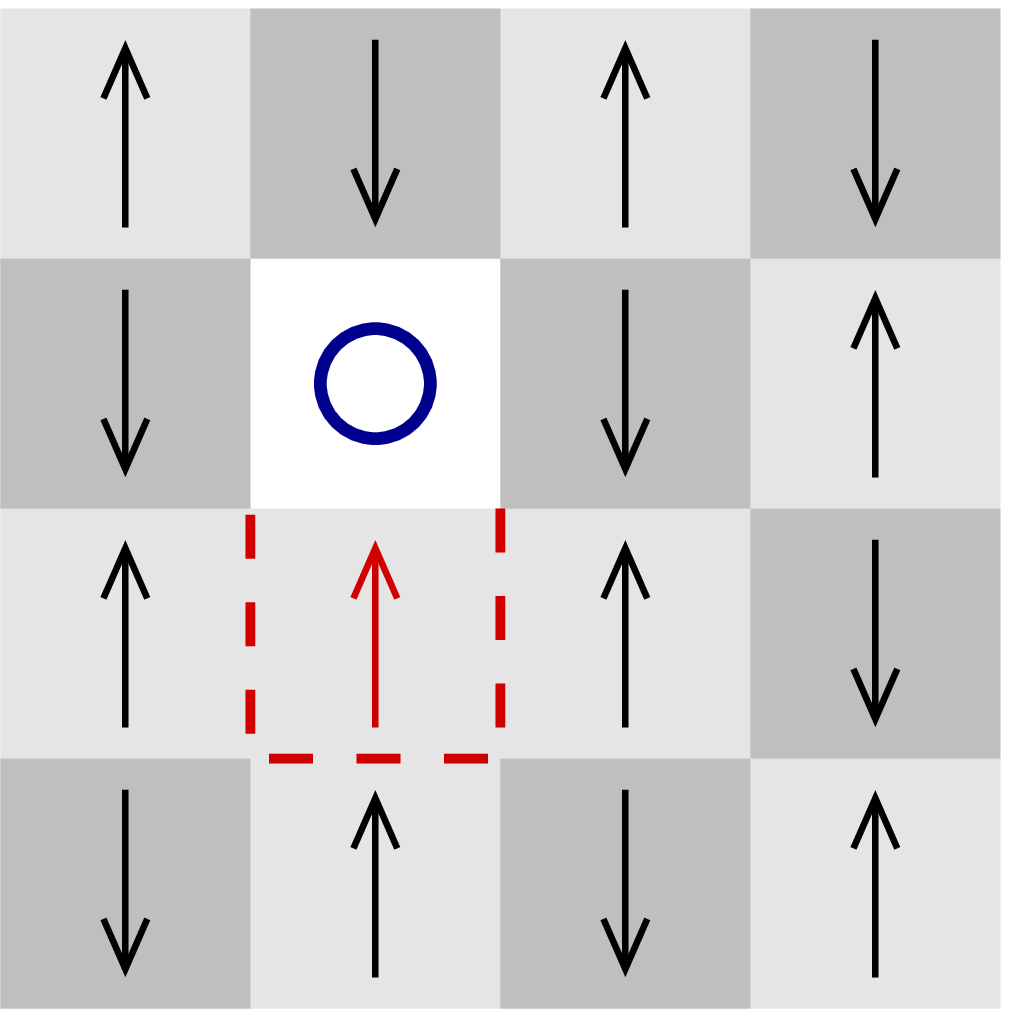}\hspace{0.005\textwidth}
\includegraphics[width=0.13\textwidth]{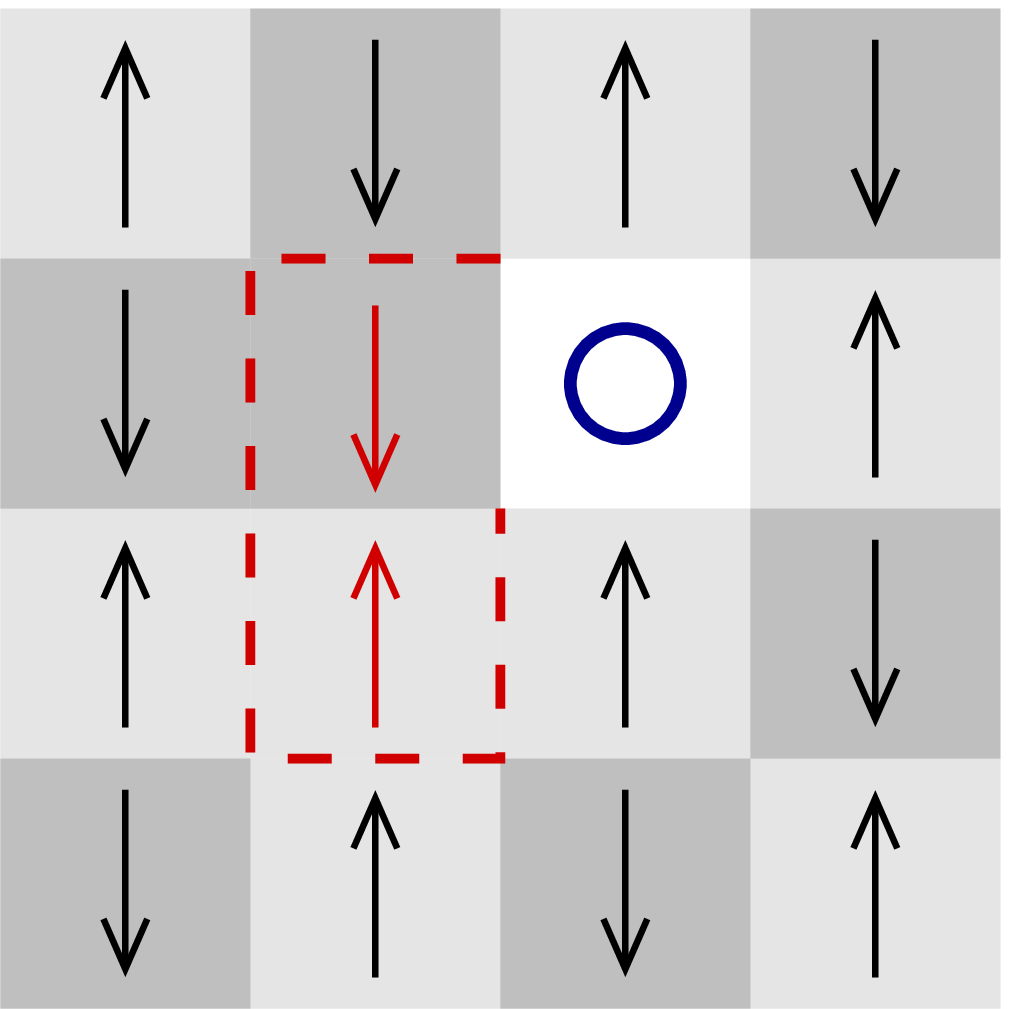}\hspace{0.005\textwidth}
\includegraphics[width=0.13\textwidth]{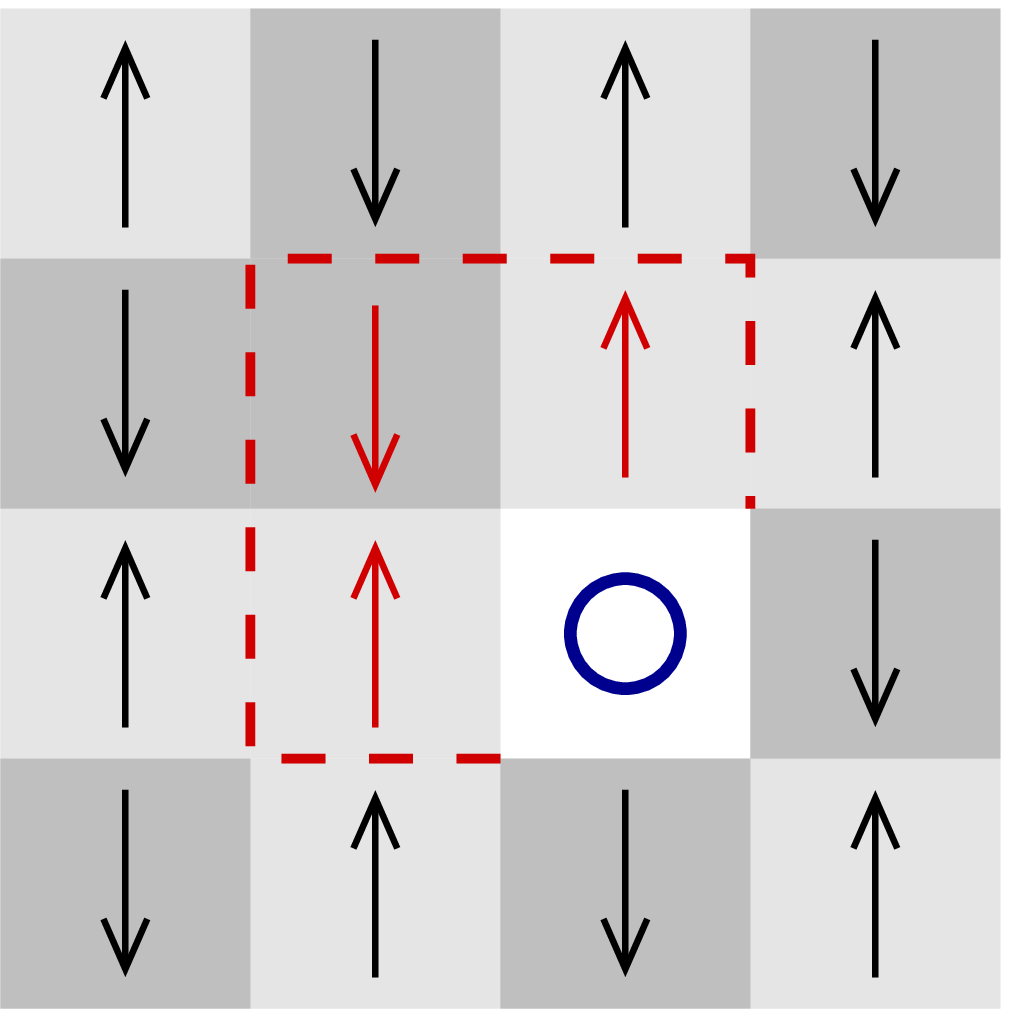}\hspace{0.005\textwidth}
\includegraphics[width=0.13\textwidth]{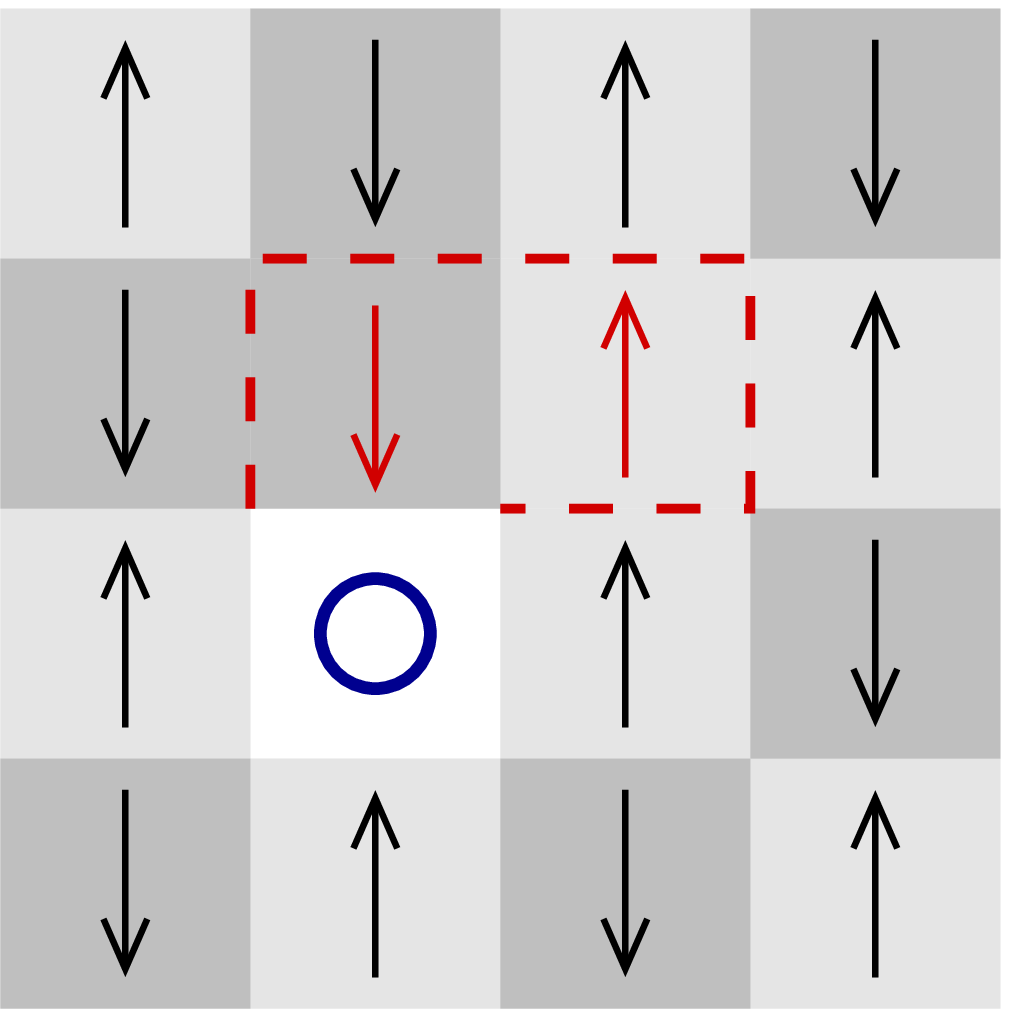}\hspace{0.005\textwidth}
\includegraphics[width=0.13\textwidth]{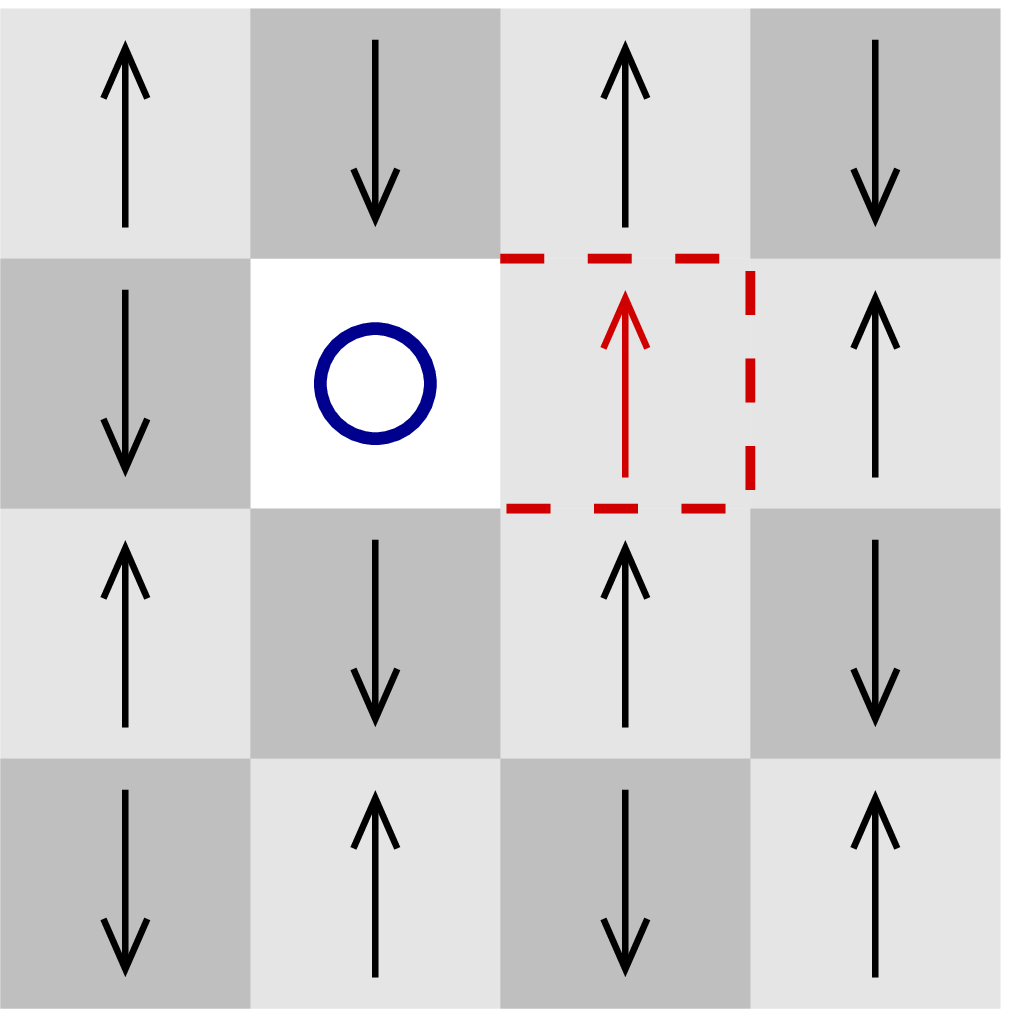}\hspace{0.005\textwidth}
\includegraphics[width=0.13\textwidth]{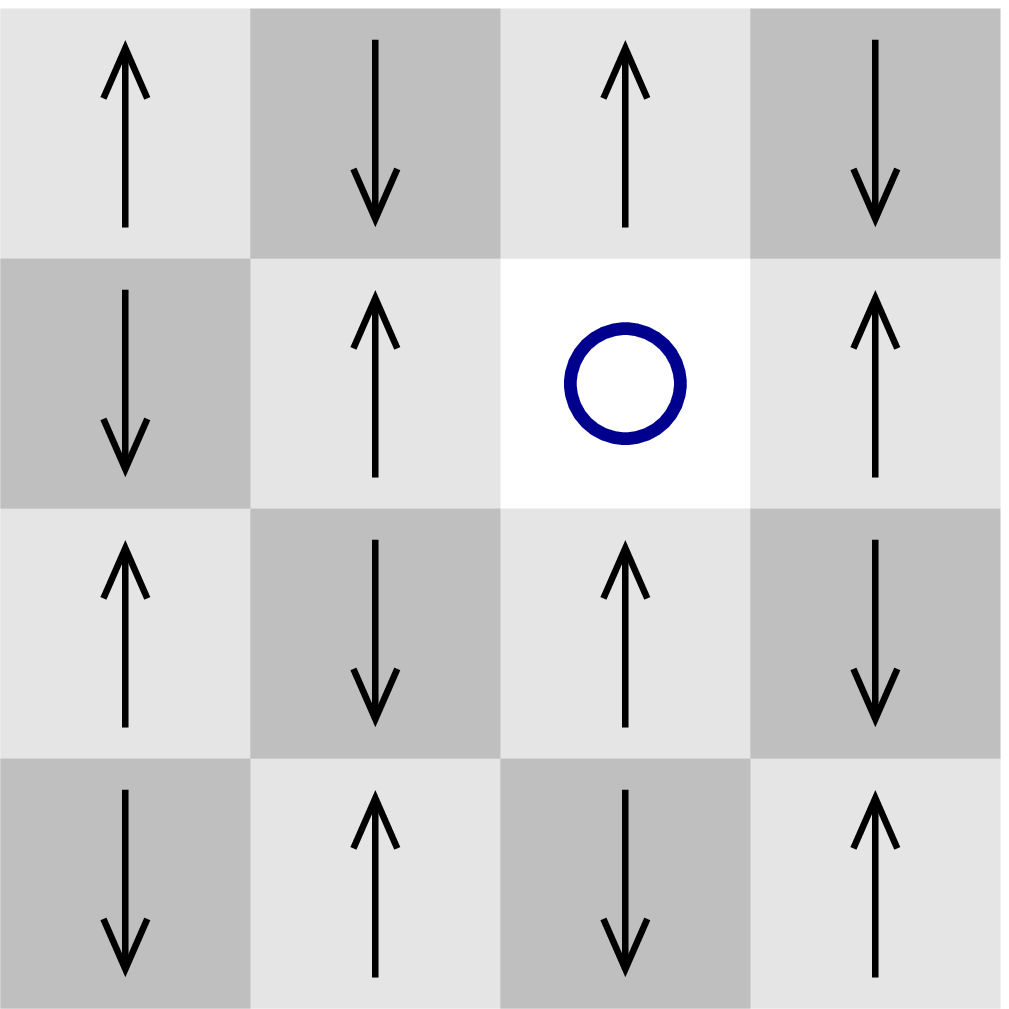}
\caption{The simplest `Trugman path' that allows for propagation of a hole in a N\'{e}el antiferromagnet~\cite{Tr88}.
After six steps, the hole has moved by two sites to a next-nearest neighbour and the N\'{e}el spin order is restored. }
\label{fig:trugpath}
\end{figure}

In the absence of quantum spin fluctuations free-particle like motion of a hole is completely suppressed.
Surprisingly enough, a hole can propagate through so-called `Trugman paths'~\cite{Tr88}, which realize a certain combinatorial rearrangement of spins compatible with the spin order.
The simplest of these processes, where the hole moves by two sites and restores the antiferromagnetic spin alignment, is shown in Fig.~\ref{fig:trugpath}.
In contrast to the situation for an almost free particle
propagation through such processes depends entirely on the successive  
rearrangement of spins in precise coordination with the hole motion, just like executing a complicated dance pattern.
With respect to the correlations between spin and hole motion we call this a correlated transport process.

The focus of the present contribution is on similar correlated transport processes in a simpler fermion-boson model. 
This model is introduced in the next section (Sec.~\ref{sec:model}).
In Sec.~\ref{sec:drude} we define the Drude weight as the appropriate measure of the efficiency of a correlated transport process.
In Sec.~\ref{sec:3bosonGF} we derive the one-fermion Green's function, which is discussed together with the Drude weight in Sec.~\ref{sec:GFandD},
before we conclude in Sec.~\ref{sec:conclusions}.

\section{The model}\label{sec:model}
The picture sketched here --- how the motion of a particle depends on the creation of background distortions and their decay --- is very general, and applies e.g. to the polaron problem of electron motion in a deformable lattice studied using the Holstein model~\cite{Ho59a,Ho59b}, and to charge transport in the presence of strong electronic correlations, such as for hole-doped Mott insulators studied using the t-J-model~\cite{BR70,KLR89}.
For that reason, we recently introduced a simple but general fermion-boson model that captures the fundamental aspects without modelling any very specific details.  
This model was originally proposed by Edwards~\cite{Ed06} and subsequently used to study single-particle quantum transport~\cite{AEF07} and metal-insulator transitions~\cite{WFAE08,EHF09,EF09}.

\begin{figure}
\centering
\includegraphics[width=0.8\textwidth]{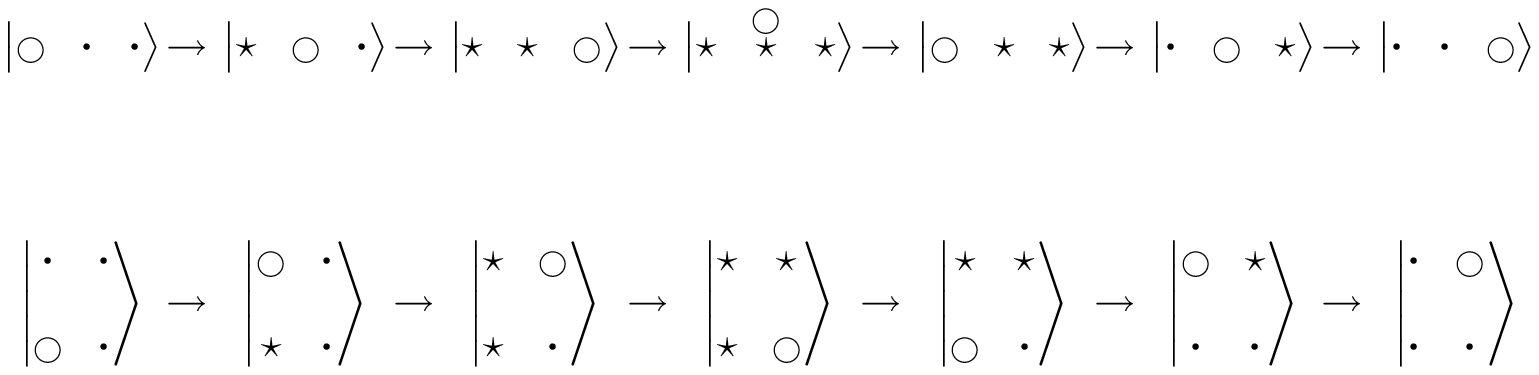}
\caption{Lowest order vacuum-restoring process that allows for propagation of a fermion in the model Eq.~\eqref{Ham}, on a 1D chain (upper row) and 2D square lattice (lower row).
In any dimension, it comprises six steps and propagates the fermion by two sites.
In steps 1--3, three bosons are excited, which are consumed in steps 4--6.
The efficiency of this process consequently scales as $t_b^6/\omega_0^5$ (see Sec.~\ref{sec:3bosonDrude}).
Note that the 2D process is the exact translation of the Trugman path shown in Fig.~\ref{fig:trugpath},
where the correlated many-particle background of the spin model is replaced by the bosonic vacuum.
For our model, such processes play a role already in 1D.
}
\label{FigEchter}
\end{figure}

We here consider the Hamiltonian
\begin{equation}\label{Ham}
H =  -t_b \sum_{\langle i, j \rangle}  c_j^{\dagger} c_i (b_i^{\dagger}
+ b^{}_j)  \\
+ \omega_0 \sum_i b_i^{\dagger} b_i % + \frac{N \lambda^2}{\omega_0} \;.
\end{equation}
for a spinless fermion (created by $c^\dagger_i$) coupled
to bosons (created by $b^\dagger_i$) of energy $\omega_0$. 
The first term in the Hamiltonian specifies how bosons are
created (and annihilated) when the fermion hops to a nearest neighbour.
 Any hop of the fermion changes the number of bosons by one. In this sense, fermion motion is boson-controlled.

Fermion hopping creates a boson only on the site the fermion leaves, and destroys a boson only on the site the fermion enters.
As a consequence motion of the fermion creates, in a lattice of arbitrary dimension, a `string' of bosonic excitations:
In each hop the fermion creates additional bosons that can be destroyed only by returning to sites already visited.
This `string effect' restricts propagation just as for the hole in the 2D antiferromagnet.

The original model (in Refs.~\cite{Ed06,AEF07,WFAE08}) contains an additional term $- \lambda \sum_i (b_i^{\dagger} + b_i)$ that accounts for boson relaxation. 
Bosons can decay when $\lambda \ne 0$, so that in this case the string effect is weakened.
In the present situation, the string effect is not overcome by boson relaxation. 
But just as for the Trugman paths, there exist processes whereby the fermion propagates, while the boson vacuum is restored~\cite{AEF07}.
The simplest of these vacuum-restoring processes comprises
six steps, promoting the fermion by two sites (see Fig.~\ref{FigEchter}).
If we compare to Fig.~\ref{fig:trugpath}, we see
that the 2D process in our model is an exact
representation of a Trugman path.
Here, such processes occur already in 1D, and can be considered as a
projection of the corresponding 2D process.
In our model such correlated transport processes emerge quite naturally from some basic assumptions. 
Our general model now allows us to study the implications of the existence of such processes for the motion of a fermionic particle.
An analytical calculation of the one-fermion Green's function, which is only possible in the $\lambda=0$-case considered here, enables us to characterize correlated transport using the Drude weight.
We restrict ourselves to the study of a single fermion in 1D at zero temperature.

\section{Drude weight}\label{sec:drude}

Transport of almost free particles is conveniently characterized through the particle mass and an effective mean free path.
How, then, can transport be characterized in a situation where it differs entirely from free particle motion?
As stressed by Kohn~\cite{Ko64} one should consider the Drude weight $D$,
which can be defined through its relation 
\begin{equation}\label{Drude}
 D= 1/(2m^*) \;, \quad \frac{1}{m^*} = \frac{\partial^2 E(k)}{\partial k^2}\Big|_{k=0} 
\end{equation}
to the quasiparticle mass $m^*$, where $E(k)$ is the quasiparticle dispersion.
Alternatively, the Drude weight can be introduced through the relation $\Re \sigma(\omega) = 2 \pi D \,\delta(\omega) + \sigma_\mathrm{reg}(\omega)$ for the real part of the conductivity $\sigma(\omega)$,
or be obtained from the f-sum rule, which relates the integrated conductivity to the kinetic energy.
Note that for our model with boson-controlled hopping, the current operator is more complicated than usual~\cite{AEF07}.

The Drude weight generalizes the original Drude theory with its concept of an effective mean free path between two scattering events, which is valid only for an almost free particle.
Note that just through the above relation between Drude weight and quasiparticle mass the latter is identified as a relevant quantity in a rigorous quantum transport theory.
Indeed, the quasiparticle mass itself is a rather complicated quantity, which can be  
strongly renormalized as a consequence of correlated many-particle dynamics.

\section{Derivation of the Green's function with a 3-boson constraint}\label{sec:3bosonGF}

\begin{figure}
\centering
\includegraphics[width=0.48\textwidth]{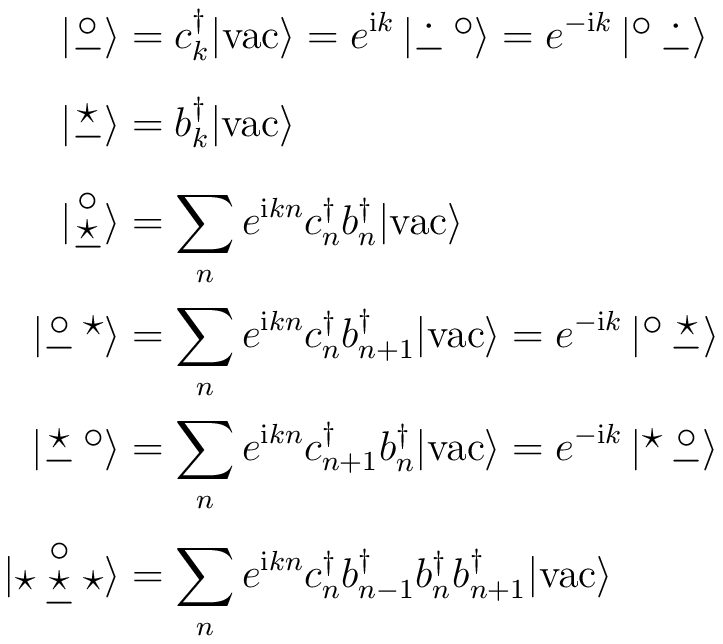}
\hspace*{1ex}
\begin{minipage}[b]{0.48\textwidth}
\caption{\label{fig:GraphicalStates}
  Examples of the graphical notation for fermion-boson states.
  An empty site is represented by a dot ($\cdot$), the fermion by a circle ($\circ$), and a boson by a star ($\star$).
  The origin is indicated with an underscore (\_).
  All states are Bloch states with momentum $k$.
The last row shows the symmetric state which
appears in the 6-step-process (cf. Fig.~\ref{fig:ThreeBosonTree}).}
\end{minipage}
\end{figure}

Even for the simple-looking Hamiltonian in Eq.~\eqref{Ham} an exact calculation of Green's functions is not possible.
The situation changes if an additional constraint is imposed, namely a bound on the number of simultaneously excited bosons.
Since the vacuum restoring 6-step-process in Fig.~\ref{FigEchter} involves 3 bosons, a non-trivial result is obtained allowing for at most 3 bosons in the system (we will justify this restriction further in the next section).
Then, only a finite number of states contribute to the one-fermion Green's function
\begin{equation}\label{GreenTC}
  G(k,z) = \langle\mathrm{vac}| c_k [z-H]^{-1} c^\dagger_k |\mathrm{vac}\rangle \;.
\end{equation}
This enables us to derive a closed expression for this Green's function.

\subsection{The tree of states}

\begin{figure}
\centering
\includegraphics[width=0.7\textwidth]{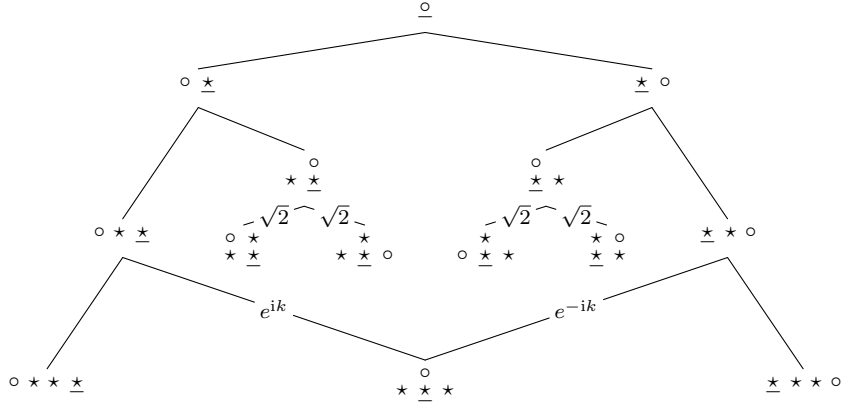}
\caption{Graph structure for the calculation of $G(k,z)$.
Each unlabelled edge corresponds to a matrix element $t_b$, 
the labelled edges to $t_b$ times the given factor.
The $\sqrt{2}$ arises from matrix elements of bosonic operators.
}
\label{fig:ThreeBosonTree}
\end{figure}

As in Fig.~\ref{FigEchter}, we use a graphical notation for states, explained in Fig.~\ref{fig:GraphicalStates}.
The calculation requires the use of Bloch states with given momentum $k$.
As an example, $|\circ\rangle = \sum_n e^{\ii k n} c^\dagger_n|\mathrm{vac}\rangle$.
The graph in Fig.~\ref{fig:ThreeBosonTree} shows all states with at most $3$
bosons that are accessible from $|\circ\rangle$ by repeated
application of $H$,
while the edges of the graph correspond to non-zero matrix elements of $H$.

The graph of states contains a loop corresponding to the 6-step-process.
To obtain a simple expression for the Green's function with a recursive calculation we must get rid of this loop.
To this end we perform a change of basis that exploits the reflection symmetry of the Hamiltonian~\eqref{Ham}, and rearrange all states in the tree structure shown in Fig.~\ref{fig:ThreeBosonTree2}.
Note that the states in this tree are not
eigenstates of the parity (or reflection) operator, since they all
have finite momentum $k$.
Note further that it is important to choose the position of the origin correctly. The origin in the two states below the tree root is shifted by one site.
Then the symmetric state in the last line of Fig.~\ref{fig:GraphicalStates}, which appears at the
bottom of the loop in Fig.~\ref{fig:ThreeBosonTree}, appears only in
the left subtree. The corresponding linear combination with a
minus sign, which would appear in the right subtree, is zero due to reflection symmetry.
Therefore Fig.~\ref{fig:ThreeBosonTree2} has no loop. 
The shift of the origin accounts for the two sites the fermion
propagates in the 6-step-process.

\subsection{Recursive calculation of the Green's function}
\begin{figure}
\centering
\includegraphics[width=\textwidth]{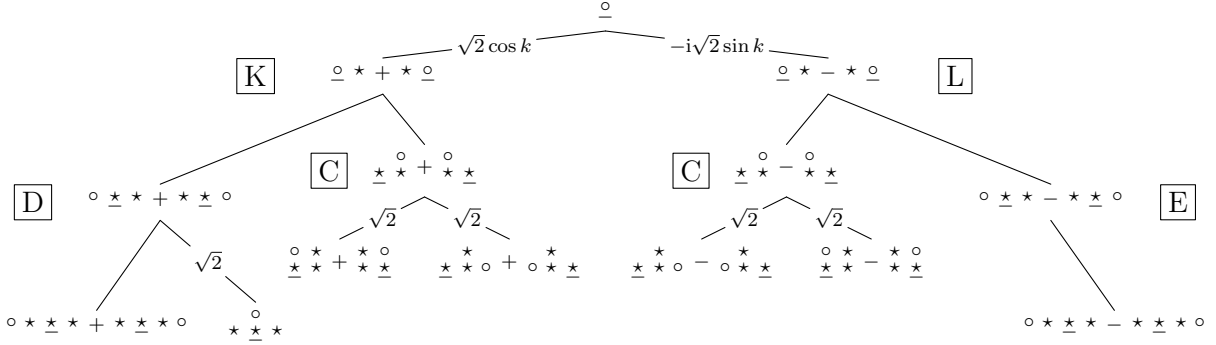}
\caption{Rearranged tree structure for the 
  calculation of $G(k,z)$ after a basis change.
  Note that the matrix element $\sqrt{2} t_b$ at the edge in the D-subtree 
  results from the scalar product of the two corresponding states,
  and not from bosonic operators. }
\label{fig:ThreeBosonTree2}
\end{figure}

The tree in Fig.~\ref{fig:ThreeBosonTree2} allows for 
the recursive calculation of $G(k,z)$.
The basic elements in this calculation are Green's functions $G_s(z)$ associated to a state $|s\rangle$ and the subtree with $|s\rangle$ at its top.
Such a Green's function is defined through the usual equation $G_s(z)= \langle s| [z-H_s]^{-1} |s\rangle$ (similar to Eq.~\eqref{GreenTC}), where $H_s$ is the projection of the Hamiltonian onto the subspace spanned by all states in the subtree.
Note that these projected Hamiltonians commute for disjoint subtrees, so that the  corresponding Green's functions can be calculated independently. 
It is for this reason that we had to remove the loop in the original graph of states (Fig.~\ref{fig:ThreeBosonTree}).

At each step of the calculation, a Green's function $G_s(z)$ associated to a subtree is written in the form
\begin{equation}\label{Greens}
G_s(z)=[z- a - b_l^2 G_l(z) - b_r^2 G_r(z)]^{-1} \;,
\end{equation}
where $G_l(z)$ and $G_r(z)$ are the Green's functions associated to the left and right subtrees one step further down the tree, $b_l$ and $b_r$ are the matrix elements of $H$ along the corresponding edges,
and $a=\langle s|H|s\rangle$ is the diagonal matrix element of $H$ in the state $|s\rangle$.
Note that when $|s\rangle$ is at the bottom of the tree, the associated  Green's function is simply given by $G_s(z)=[z-a]^{-1}$.

To justify Eq.~\eqref{Greens} we proceed in a standard way and split $H_s$ as
\begin{equation}
 H_s = a |s\rangle\langle s| + b_l  V_{l}+ H_l + b_r V_r +  H_r \;,
\end{equation}
where $|l\rangle$, $|r\rangle$ correspond to the states associated with $G_l(z)$, $G_r(z)$,
and $V_{l} = |s\rangle\langle l| + |l\rangle\langle s|$, $V_{r}= |s\rangle\langle r| + |r\rangle\langle s|$.
Multiplying both sides of the operator identity
\begin{equation}
1 = [z-H_r - H_l]^{-1} [z-H_s \, + \, a |s\rangle\langle s| + b_l V_l + b_r V_r] 
\end{equation}
with $[z-H_s]^{-1}$ from the right, we get the (generalized) Dyson equation
\begin{equation}
[z-H_s]^{-1} = [z-H_r - H_l]^{-1}  + [z-H_r - H_l]^{-1} \, [a |s\rangle\langle s| + b_l V_l + b_r V_r] \,  [z-H_s]^{-1} \;.
\end{equation}
Taking scalar products with the state $|s\rangle$ leads to 
\begin{equation}\label{Gs1}
 G_s(z) = z^{-1} + z^{-1} \Big( a G_s(z) + b_l \langle l| [z-H_s]^{-1} |s\rangle + b_r \langle r| [z-H_s]^{-1} |s\rangle \Big)  \;, 
\end{equation}
and in a second step, with $|l\rangle$, $|r\rangle$, to
\begin{equation}\label{Gs2}
\begin{split}
 \langle l| [z-H_s]^{-1} |s\rangle &= G_{l} b_{l} G_s(z) \;, \\
  \langle r| [z-H_s]^{-1} |s\rangle &= G_{r} b_{r} G_s(z) \;.
 \end{split}
\end{equation}
Inserting Eq.~\eqref{Gs2} into Eq.~\eqref{Gs1} and solving for $G(s)$ yields our result Eq.~\eqref{Greens}.

We now apply Eq.~\eqref{Greens} to the tree of states in Fig.~\ref{fig:ThreeBosonTree2}.
Working from the top of the tree of states to the bottom, one obtains 
\begin{equation}\label{Green3}
  \begin{split}
    G(k,z) & = \Big[z - 2t_b^2(K(z) \cos^2 k + L(z) \sin^2 k ) \Big]^{-1} \\
    &=  \Big[z - t_b^2 (K(z)+L(z)) - t_b^2 (K(z)-L(z)) \cos 2k \Big]^{-1} \;,
  \end{split} 
\end{equation}
where the intermediate results to the subtrees labelled in
Fig.~\ref{fig:ThreeBosonTree2} are
\begin{equation}
  \begin{split}
    C(z) &= \Big[z-2\omega_0 - \frac{4t_b^2}{z-3\omega_0}
    \Big]^{-1} \;,\\
    D(z) &= \Big[z-2\omega_0 - \frac{3t_b^2}{z-3\omega_0}
    \Big]^{-1} \;,\\
    E(z) &= \Big[z-2\omega_0 - \frac{t_b^2}{z-3\omega_0}
    \Big]^{-1} \;,\\
    K(z) &= \Big[z-\omega_0 - t_b^2(C(z)+D(z)) \Big]^{-1} \;,\\
    L(z) &= \Big[z-\omega_0 - t_b^2(C(z)+E(z)) \Big]^{-1}  \;.
  \end{split}
\end{equation}

\subsection{Discussion}

Our derivation of the Green's function is closely related to well-known methods such as the renormalized perturbation expansion~\cite{Ec83}, the (Lanczos) recursion method~\cite{HHK72} or the Mori-Zwanzig formalism~\cite{Fu91}.
For our problem the most important aspect is the appearance of two Green's function in Eq.~\eqref{Greens}, corresponding to the two branches of each node in the binary tree of states in Fig.~\ref{fig:ThreeBosonTree2}.
This generalizes the standard recursive construction leading, e.g. in Lanczos recursion, to a continued fraction of the form
\begin{equation}\label{GreenCF}
  G(z) = \cfrac{1}{z- a_0 -
    \cfrac{b_1^2}{z-a_1-\cfrac{b_2^2}{z-a_2-\dots}}} \;.
\end{equation}
In the present case this construction would result in very complicated expressions, since arbitrary linear combinations of the states shown in Figs.~\ref{fig:ThreeBosonTree},~\ref{fig:ThreeBosonTree2} occur in the recursion, e.g. of states with different boson number.
The additional freedom provided by Eq.~\eqref{Greens} is important to obtain simple results for $G(k,z)$ in closed form (obviously, our result is not given in the form of Eq.~\eqref{GreenCF}).
The success depends on our ability to find a suitable arrangement of states in a tree as in Fig.~\ref{fig:ThreeBosonTree2}. 
This task can hardly be automatized, and such an arrangement is difficult to find for more complicated situations. Nevertheless, the idea presented here can be generalized so that it may be useful also in further calculations.

\section{Evaluation of the Green's function and Drude weight}\label{sec:GFandD}

\begin{figure}
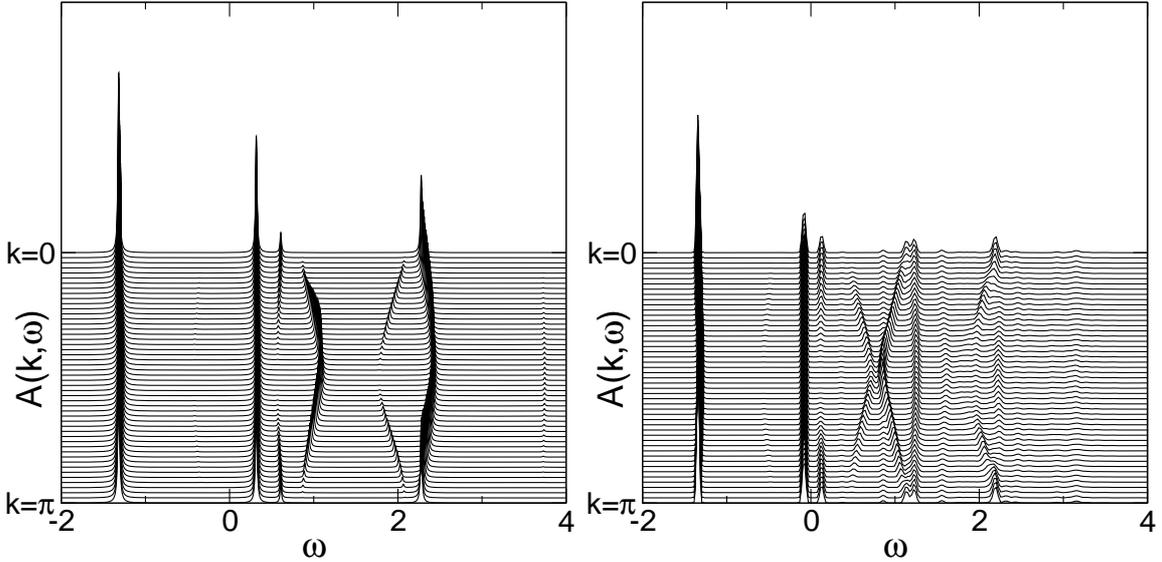

\centering
\includegraphics[width=0.47\textwidth]{fig7a.eps}
\includegraphics[width=0.47\textwidth]{fig7b.eps}
\caption{Spectral function $A(k,\omega)$ for $t_b=1.0$ and
  $\omega_0 = 1.0 $ with the 3-boson constraint from Eq.~\eqref{Green3}
  (left panel) compared to the numerical result (right panel, from Ref.~\cite{AEF07}) without the constraint.}
\label{fig:Spectral3}
\end{figure}

In Fig.~\ref{fig:Spectral3} we show the spectral function 
\begin{equation}
A(k,\omega)= - (1/\pi) \Im G(k,z) \;.
\end{equation}
We see that the Green's function is periodic in $k$ with period $\pi$.
The reason is that the vacuum-restoring 6-step process (Fig.~\ref{FigEchter}) propagates the fermion by two sites, resulting in the term $\propto \cos 2k$ in Eq.~\eqref{Green3}.
The spectral function is dominated by a coherent quasiparticle band at low energy, which is almost dispersionless and separated by a large gap (of the order $\omega_0$) from other (coherent) signatures at higher energies. Note that the dispersion of the quasiparticle band is related to the Drude weight, as discussed in the next subsection. 
As Fig.~\ref{fig:Spectral3} further shows, $G(k,z)$
compares quite well to the numerical result without the 3-boson-constraint~\cite{AEF07}.
The reason is that, for all but very small $\omega_0 / t_b \simeq 1$, 
hole motion is dominated by the 6-step process that involves only 3 bosons.
We will next discuss how this fact shows up in the Drude weight.

\subsection{The Drude weight}\label{sec:3bosonDrude}

\begin{figure}
\centering
\includegraphics[width=0.45\textwidth]{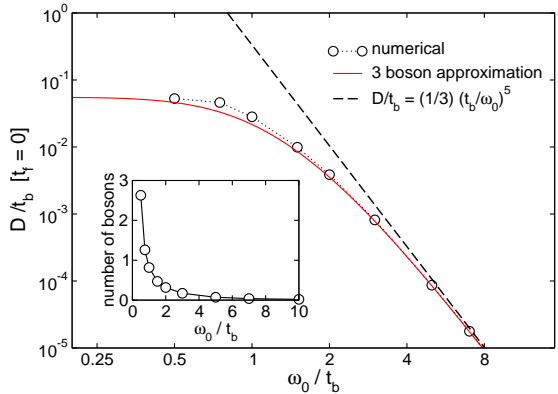}
\hspace*{2ex}
\begin{minipage}[b]{0.48\textwidth}
\caption{\label{fig:Drude3}
 Drude weight $D$ with 3-boson constraint, and from exact
  numerics without the constraint~\cite{AEF07}.
The dashed line gives the asymptotic result $D = t_b^6/(3\omega_0^5)$ for large $\omega_0/t_b$.
The inset displays the average number of bosons in the ground-state without the 3-boson constraint.}
\vspace*{1ex}
\end{minipage}
\end{figure}

Following Eq.~\eqref{Drude} the Drude weight $D$
is obtained from the ground-state energy 
\begin{equation}
E(k) = E_0 + (D/2) k^2 + O(k^4) \;,
\end{equation}
which is given by lowest pole of $G(k,z)$.
With decreasing $\omega_0$, the Drude weight grows as boson-controlled transport becomes more efficient, but remains finite for $\omega_0 \to 0$ (see Fig.~\ref{fig:Drude3}).
Since the 6-step-process involves 5 intermediate states with excited
bosons, the leading order of $D/t_b$ in an expansion in $t_b/\omega_0$ is $(t_b/\omega_0)^5$.
Any process contributing to $D/t_b$ that has an intermediate state
with more than 3 bosons, must be of order $(t_b/\omega_0)^7$ or higher
since at least $4$ bosons have to be excited and de-excited afterwards,
which requires $8$ steps.
Therefore the asymptotic behaviour of $D$ for large $\omega_0/t_b$  can be obtained from $G(k,z)$ in Eq.~\eqref{Green3}, with the result $D \simeq  t_b^6/(3 \omega_0^5)$ shown in Fig.~\ref{fig:Drude3}.

Once again, we compare our analytical $D$ to the numerical result without the 3-boson constraint.
Both results fit surprisingly well
although the restriction to a small number of bosons might appear to be a crude approximation for small $\omega_0/t_b$.
But the inset in Fig.~\ref{fig:Drude3} shows that in numerical calculations without the 3-boson constraint the number of bosons in the ground-state remains small over a large range of values of $\omega_0/t_b$.
This indicates, together with the agreement between the two curves for $D$, that the main contribution to fermion propagation comes from the 6-step-process even when $\omega_0/t_b$ is small, although higher order vacuum-restoring processes involving more than three bosons might become energetically favourable in this case.
One explanation for that behaviour is that correlated transport through vacuum-restoring processes relies on the correlated excitation and de-excitation of bosons.
Higher order processes involving more than 3 bosons
cannot give a large contribution to $D$ since correlations between many bosons are inevitably reduced.
This also explains the saturation of $D$ as $\omega_0 \to 0$:
The contribution of the 6-step-process to $D$ remains finite for $\omega_0 \to 0$ (restricting to 3 bosons,
the Hamiltonian is a bounded operator, and $E(k)$ and $D$ remain finite).
Then, if the main contribution to $D$ comes from the 6-step-process,
$D$ itself remains finite for $\omega_0\to 0$ even without the 3-boson constraint.

\section{Conclusions and Outlook}\label{sec:conclusions}

In this contribution we discuss correlated transport
through vacuum-restoring processes in a general fermion-boson model.
We obtain a simple analytical expression for the one-fermion Green's function. 
>From the Green's function we deduce the Drude weight which characterizes transport even when the description in terms of a mean-free path between two scattering events, as in Drude's original theory, breaks down. 
The dependence of the Drude weight on the boson energy $\omega_0/t_b$ reveals two opposing tendencies in the correlated transport process:
On the one hand, transport is assisted by bosons, and therefore $D$ increases with decreasing $\omega_0/t_b$.
On the other hand, it is the involvement of bosonic excitation in the transport process itself that limits its efficiency, and $D$ remains finite in the limit of small $\omega_0/t_b$. 
This competition between correlations and fluctuations is a general physical phenomenon, which can be further studied in the full model~\cite{AEF07}.

The correlated transport process studied here is also important at finite fermion density, when the formation of a charge-density wave at half-filling triggers a metal-insulator transition~\cite{WFAE08}.
The charge-density wave insulator is peculiar in that it involves few bosons, in contrast to the common Peierls scenario, and is caused by strong boson-fermion correlations. A hole doped into this insulator propagates through the 6-step-process discussed here, while an additional electron can propagate through a more efficient 2-step-process.    
The simple results obtained here for a single fermion are important to understand the more complicated physics in this case. 
Our considerations may also allow for the development of an analytical description of the correlated charge-density wave insulator. 

\ack
The authors are grateful for discussions with S. Ejima, T. Koch, and G. Wellein.
This work was supported by Deutsche Forschungsgemeinschaft
through SFB 652.

\section*{References}

\providecommand{\newblock}{}

\end{document}